\documentclass[aps,twocolumn,floatfix,groupedaddress,nofootinbib]{revtex4}
\usepackage{csquotes}
\usepackage{graphicx,color}
\usepackage{float}
\usepackage[caption=false]{subfig}
\usepackage{amsmath}
\usepackage{amsmath}
\usepackage{multirow}
\usepackage{dsfont}
\usepackage[shortlabels]{enumitem}

\begin{document}

\title{Description of longitudinal modes in moderately coupled Yukawa systems with the static local field correction\vspace*{-3.50mm}}
\author{P. Tolias$^{*}$ and F. Lucco Castello}
\affiliation{Space and Plasma Physics, Royal Institute of Technology, Stockholm, SE-100 44, Sweden\\
             $^{*}$\,Author to whom correspondence should be addressed: tolias@kth.se\vspace*{-2.80mm}}
\begin{abstract}
\noindent In moderately coupled Yukawa fluids, longitudinal mode dispersion is determined by the competition between kinetic and potential effects. In a recent paper [Khrapak and Cou{\"e}del, Phys. Rev. E  102, 033207 (2020)], a semi-phenomenological dispersion relation was constructed by the ad-hoc addition of the Bohm-Gross kinetic term to the generalized instantaneous excess bulk modulus, which showed very good agreement with simulations. In this paper, a nearly identical dispersion relation is derived in a rigorous manner based on a dielectric formulation with static local field corrections. At moderate coupling, this formalism is revealed to be more accurate than other successful theoretical approaches.
\end{abstract}
\maketitle

\noindent \emph{Yukawa one-component plasmas (YOCP)} are model systems that consist of charged point particles which are immersed in a polarizable charge-neutralizing background. The particle interactions follow the Yukawa pair potential $U(r)=({Q^2}/{r})\exp{\left(-{r}/{\lambda}\right)}$ with $Q$ the particle charge and $\lambda$ the screening length that is determined by the polarizable medium\,\cite{dusrev1,dusrev2,dusrev3}. Thermodynamic states are typically specified in terms of the coupling parameter $\Gamma=\beta{Q}^2/d$ and the screening parameter $\kappa=d/\lambda$, with $\beta=1/(k_{\mathrm{B}}T)$ the inverse temperature, $d=(4\pi{n}/3)^{-1/3}$ the Wigner-Seitz radius and $n$ the particle density. Due to the YOCP softness spanning from long range bare Coulomb ($\kappa\to0$, OCP) to short range hard sphere interactions ($\kappa\to\infty$) and its relevance to various strongly coupled laboratory systems\,\cite{dusexp1,dusexp2,dusexp3}, its thermodynamic, structural, phase and dynamic properties have been exhaustively studied\,\cite{dusrev1,dusrev2,dusrev3}.

Collective modes in three-dimensional YOCP systems have been investigated with various theoretical methods; phenomenological approaches with restricted microscopic basis such as generalized hydrodynamics\,\cite{theoGH1,theoGH2,theoGH3}, viscoelastic dynamic density functional hydrodynamics\,\cite{theoVD1,theoVD2} or memory function formalisms\,\cite{theome1,theome2}, first-principle microscopic approaches such as the quasi-localized charge approximation (QLCA)\,\cite{theoQL1,theoQL2,theoQL3} or the dielectric formulation with local field corrections\,\cite{theosl1,theosl2} and pure mathematical models constructed to obey exact self-consistency conditions such as the method of moments (MOM)\,\cite{theosr1,theosr2,theosr3}. Moreover, attention has been paid to the long wavelength limit and the acquisition of practical expressions for the sound speed\,\cite{theocs1,theocs2,theocs3,theocs4}. The ever-growing availability of exact results from molecular dynamics (MD) simulations can assist to distinguish limitations, quantify the level of accuracy and determine the different applicability ranges of these theoretical methods\,\cite{theoVD1,theome2,theocs2,theocs3,MDwave1,MDwave2,MDwave3,MDwave4,MDwave5,modulig}.

In the last two decades, YOCP wave investigations focused mainly on strong coupling, \emph{i.e.} near the liquid-solid phase transition. Hence, theoretical formalisms known to perform excellently in this regime (such as the QLCA and the MOM) have gained considerable traction. Recent interest on collective modes across coupling regimes\,\cite{theocs3,modulig} calls for the revisiting of often dismissed formalisms such as the local field corrected dielectric formulation\,\cite{theosl1,theosl2}.

In this Brief Communication, we focus on the longitudinal mode at moderate coupling. In the liquid portion of the Yukawa phase diagram, this regime is bounded above by the Frenkel line (dynamic crossover corresponding to the emergence of shear stiffness at undamped natural frequencies or the appearance of the transverse mode in the spectrum)\,\cite{frenkel} and bounded below by the Kirkwood line (structural crossover from monotonic decay to exponentially damped oscillatory decay of the total correlation function)\,\cite{kirkwoo}. A recent work by Khrapak \& Cou{\"e}del\,\cite{modulig} has investigated the dispersion relations of Yukawa fluids in this regime, where kinetic and potential contributions are comparable. A semi-phenomenological dispersion relation was constructed that is based on the ad hoc addition of the Bohm-Gross kinetic term to a QLCA-inspired potential term which demonstrated very good agreement with simulation results. In what follows, we shall rigorously derive a nearly identical dispersion relation applying the static local field corrected dielectric formulation.

In the framework of the \emph{linear density-density response theory}, it can be shown that the exact density response and dielectric response functions of one-component pair-interacting systems can always be expressed as\,\cite{slfcge1}
\begin{align}
\chi(\boldsymbol{k},\omega)=\frac{\chi_0(\boldsymbol{k},\omega)}{1-U(\boldsymbol{k})\left[1-G(\boldsymbol{k},\omega)\right]\chi_0(\boldsymbol{k},\omega)}\,,\,\,\,\,\,\,\,\label{densityresponseDLFC}\\
\epsilon^{-1}(\boldsymbol{k},\omega)=1+U(k)\chi(\boldsymbol{k},\omega)\,,\quad\quad\quad\quad\quad\quad\,\,\,\,\,\label{dielectricresponsegeneral}
\end{align}
with $U(\boldsymbol{k})$ the interaction potential's Fourier transform, $G(\boldsymbol{k},\omega)$ the unknown dynamic local field correction and $\chi_0(\boldsymbol{k},\omega)$ the ideal density response. In the classical case,
\begin{align}
\chi_0(\boldsymbol{k},\omega)=-\int\frac{\boldsymbol{k}}{\omega-\boldsymbol{k}\cdot\boldsymbol{v}+\imath0}\cdot\frac{\partial{f}_0(\boldsymbol{p})}{\partial\boldsymbol{p}}d^3p\,,\,\,\,\,\,\label{densityresponseideal}
\end{align}
with ${f}_0$ the equilibrium momentum distribution function. The assumption that the dominant effect of the local field correction originates from its static value leads to\,\cite{slfcge1,slfcge2}
\begin{align}
\chi(\boldsymbol{k},\omega)=\frac{\chi_0(\boldsymbol{k},\omega)}{1-U(\boldsymbol{k})\left[1-G(\boldsymbol{k})\right]\chi_0(\boldsymbol{k},\omega)}\,,\,\,\,\,\,\,\,\,\,\,\,\,\,\label{densityresponseSLFC}
\end{align}
where $G(\boldsymbol{k},\omega)\equiv{G}(\boldsymbol{k})$ is known as static local field correction (SLFC). Note that Eq.(\ref{densityresponseSLFC}) can be formally derived by complementing the first member of the Bogoliubov-Born-Green-Kirkwood-Yvon hierarchy with the closure condition $f_2(\boldsymbol{r},\boldsymbol{p},\boldsymbol{r}^{\prime},\boldsymbol{p}^{\prime},t)=f(\boldsymbol{r},\boldsymbol{p},t)f(\boldsymbol{r}^{\prime},\boldsymbol{p}^{\prime},t)Q(\boldsymbol{r}-\boldsymbol{r}^{\prime})$ with $f_2$ the two-particle distribution function, $f$ the one-particle distribution function and $Q$ an equilibrium correlation function\,\cite{slfcge3}. $G(\boldsymbol{k})\equiv0$ yields the mean field density response and $G(\boldsymbol{k})\equiv1$ the ideal density response.

The \emph{quantum version of the dielectric formulation} combines the SLFC dielectric response function with the quantum fluctuation-dissipation theorem and zero frequency moment rule (more simply the relation between the static and the dynamic structure factor)\,\cite{slfcge4,slfcge5}. Any ansatz that connects the equilibrium correlation function $Q$ with the static structure factor $S(\boldsymbol{k})$ leads to a coupled system of equations that can be solved iteratively. Different ansatzes lead to different approximations such as the Singwi-Tosi-Land-Sj{\"o}lander scheme\,\cite{slfcqu1}, the convolution approach\,\cite{slfcqu2} and the hypernetted-chain approach\,\cite{slfcqu3}. Despite the fact that correlations are treated classically and quantum effects are only included on the random phase approximation (RPA) level, the dielectric formalism has been discerned to be particularly successful in describing the dynamic, structural \& thermodynamic properties of quantum one-component plasmas\,\cite{slfcge4,slfcge5}. The numerical convergence of these self-consistent schemes is not straightforward due to the emergence of the infinite summation over the Matsubara frequencies as well as the integral connection between $G(\boldsymbol{k})$ and $S(\boldsymbol{k})$\,\cite{slfcge5,slfcqu3}.

The \emph{classical version of the dielectric formulation} combines the SLFC dielectric response function with the classical fluctuation-dissipation theorem\,\cite{slfccl1}
\begin{align}
S(\boldsymbol{k},\omega)=-\frac{1}{\pi{n}\beta}\frac{\Im\{\chi(\boldsymbol{k},\omega)\}}{\omega}\,,\quad\quad\,\,\,\,\,\label{fluctuationdissipationclassical}
\end{align}
and the zero frequency moment rule\,\cite{slfccl2}
\begin{align}
S(\boldsymbol{k})=\int_{-\infty}^{+\infty}S(\boldsymbol{k},\omega)d\omega\,.\quad\quad\quad\quad\quad\label{frequencymomentrule}
\end{align}
Here, $S(\boldsymbol{k},\omega)$ denotes the dynamic structure factor. The frequency integration is trivially performed exploiting the analytical nature of the complex valued density response, see the Kramers-Kronig causality relations. We get\,\cite{slfccl3}
\begin{align}
S(\boldsymbol{k})=-({n}\beta)^{-1}\chi(k,0)\,.\quad\quad\quad\quad\quad\,\,\,\label{kramerskronig}
\end{align}
Substituting for the SLFC density response of Eq.(\ref{densityresponseSLFC}) and using $\chi_0(\boldsymbol{k},0)=-n\beta$, which stems from the ideal density response of Eq.(\ref{densityresponseideal}) for Maxwellian distributions, we have
\begin{align}
n\beta{U}(\boldsymbol{k})\left[1-G(\boldsymbol{k})\right]={S^{-1}(\boldsymbol{k})}-1\,.\,\,\,\,\,\,\,\,\,\,\,\label{classicalSLFCbasic}
\end{align}
Combining Eqs.(\ref{densityresponseSLFC},\ref{classicalSLFCbasic}), we end up with
\begin{align}
\frac{1}{\chi(\boldsymbol{k},\omega)}=\frac{1}{\chi_0(\boldsymbol{k},\omega)}-\frac{1}{n\beta}\left[\frac{1}{S(\boldsymbol{k})}-1\right]\,.\label{densityresponseclassicalSLFC}
\end{align}
Since $S(\boldsymbol{k})$ can be introduced externally either from computer simulations or integral equation theory approaches, the dielectric response is now fully specified. In contrast to quantum systems, the SLFC dielectric formulation for classical systems does not lead to a convoluted relation between the static structure factor and density response.

\begin{figure}
	\centering
	\includegraphics[width=3.10in]{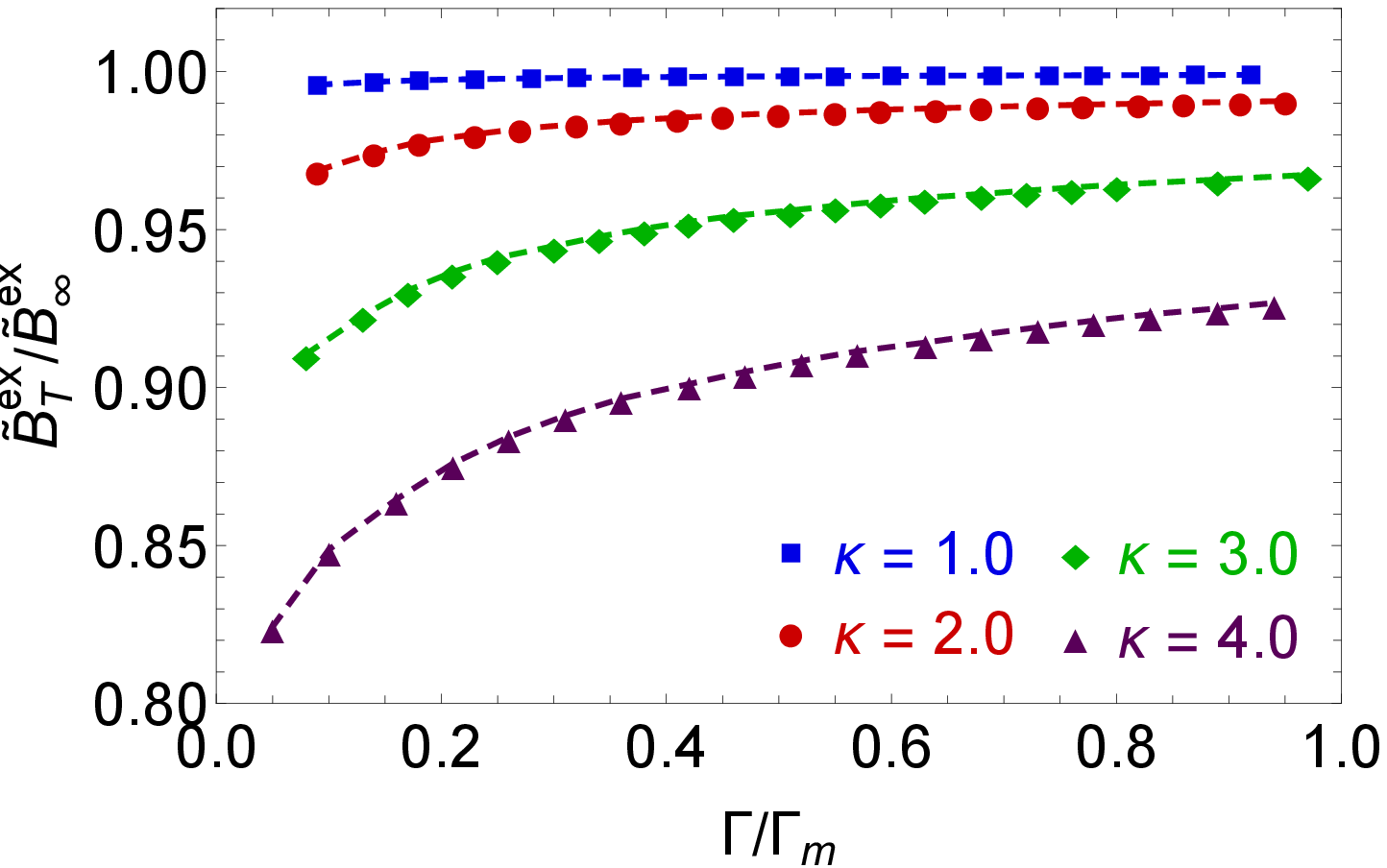}
	\caption{The quantity $\widetilde{B}^{\mathrm{ex}}_{\mathrm{T}}/\widetilde{B}^{\mathrm{ex}}_{\infty}$ as a function of $\Gamma/\Gamma_{\mathrm{m}}(\kappa)$ for $\kappa=(1,2,3,4)$. The approximate relation $\widetilde{B}^{\mathrm{ex}}_{\mathrm{T}}\simeq\widetilde{B}^{\mathrm{ex}}_{\infty}$ becomes
    inaccurate for $\kappa>3$ and breaks down for $\kappa>4$. The necessary structural, $g(x)$, and thermodynamic, $\widetilde{B}^{\mathrm{ex}}_{\mathrm{T}}$ from the virial route, input was computed with the IEMHNC approach\,\cite{IETref3}.}\label{fig:long_wavelength}
\end{figure}

\begin{figure*}
	\centering
	\includegraphics[width=6.80in]{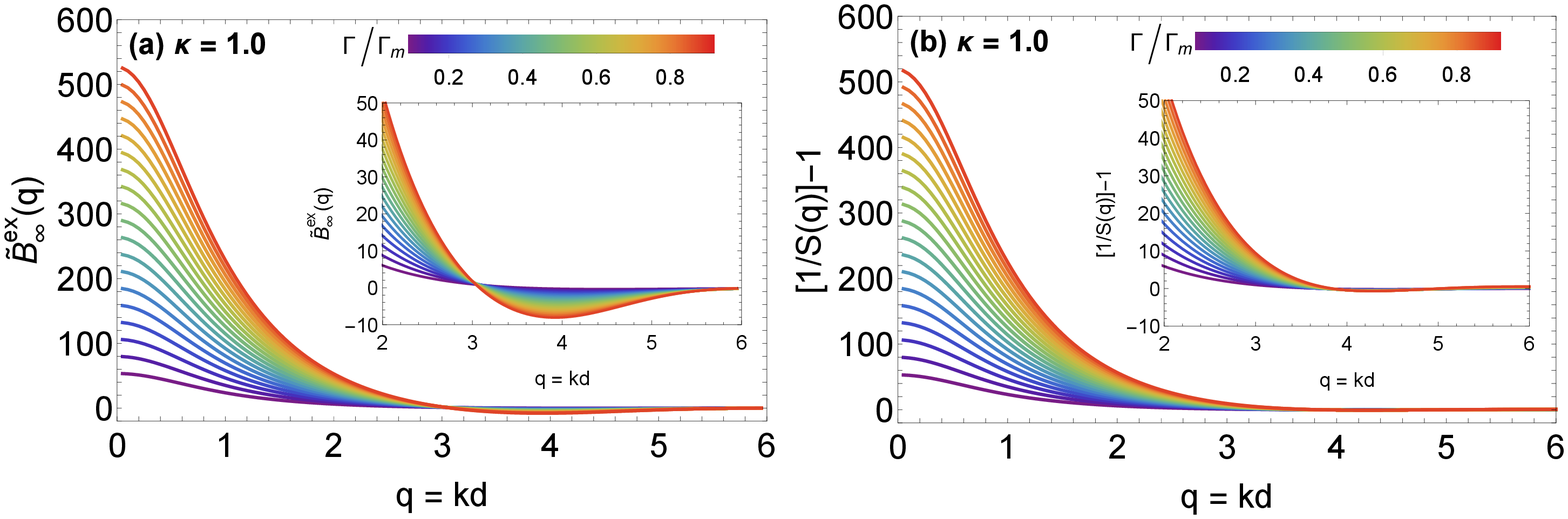}
	\caption{The quantities (a) $\widetilde{B}^{\mathrm{ex}}_{\infty}(q)$, (b) ${S}^{-1}(q)-1$, for $\kappa=1$ and varying $\Gamma/\Gamma_{\mathrm{m}}$. The structural input [$g(x),\,S(q)$] was computed with the IEMHNC approach\,\cite{IETref3}. The rough relation $\widetilde{B}^{\mathrm{ex}}_{\infty}(q)\simeq{S}^{-1}(q)-1$ breaks down at strong coupling for $q\gtrsim{3d}$\,(see inset).}\label{fig:any_wavelength}\vspace*{-3.00mm}
\end{figure*}

The study of \emph{longitudinal modes within the classical SLFC dielectric formalism} proceeds with the roots of the dielectric response; $\epsilon(\boldsymbol{k},\omega)=0$ with Eqs.(\ref{dielectricresponsegeneral},\ref{densityresponseclassicalSLFC}) leads to
\begin{align}
\left[\frac{1}{S(\boldsymbol{k})}-1\right]=\frac{n\beta}{\chi_0(\boldsymbol{k},\omega)}\,.\quad\quad\quad\quad\quad\quad\label{dispersionrelationinitial}
\end{align}
The ideal Vlasov response $\chi_0(\boldsymbol{k},\omega)$ can be expressed via the well-studied plasma dispersion function $\mathcal{Z}(\zeta)$\,\cite{slfcmo1} as
\begin{align}
&\chi_0(k,\omega)=-n\beta\left[1+\frac{\omega}{\sqrt{2}kv_{\mathrm{T}}}\mathcal{Z}\left(\frac{\omega}{\sqrt{2}kv_{\mathrm{T}}}\right)\right]\,\mathrm{where}\nonumber\\
&\mathcal{Z}(\zeta)=\frac{1}{\sqrt{\pi}}\int_{-\infty}^{+\infty}\frac{e^{-y^2}}{\zeta-y+\imath0}dy\label{plasmadispersionfunction}
\end{align}
with $v_{\mathrm{T}}=\sqrt{T/m}$ the thermal velocity. Substituting into Eq.(\ref{dispersionrelationinitial}), we get the \textbf{general SLFC dispersion relation}
\begin{align}
\left[\frac{1}{S(k)}-1\right]=-\frac{1}{1+\zeta\mathcal{Z}\left(\zeta\right)}\,\label{dispersionrelationgeneral}
\end{align}
with $\zeta=\omega/(\sqrt{2}kv_{\mathrm{T}})$.\,Such dispersion relations have been discussed by Ichimaru for classical one-component plasmas\,\cite{slfcmo2,slfcmo3} and Murillo for Yukawa one-component plasmas\,\cite{theosl1,theosl2}. Since longitudinal waves propagate in the regime of negligible Landau damping, $\omega\gg{k}v_{\mathrm{T}}$ or $\zeta\gg1$, the asymptotic expansion of the plasma dispersion function can be used. Keeping the three leading order terms, we have
$\zeta\mathcal{Z}(\zeta)=-1-(1/2)\zeta^{-2}-(3/4)\zeta^{-4}$. Performing an additional Taylor expansion, where the two leading order terms are retained, we get $[1+\zeta\mathcal{Z}(\zeta)]^{-1}=-2\zeta^2+3$. This leads to the \textbf{analytic SLFC dispersion relation}
\begin{align}
\omega^2=3k^2v_{\mathrm{T}}^2+\left[\frac{1}{S(k)}-1\right]k^2v_{\mathrm{T}}^2\,.\label{dispersionrelationclosed}
\end{align}
It should be emphasized that no substitution for the pair potential has been performed yet and that no assumption for the nature of the interactions (e.g charged or neutral) has been invoked so far. In other words, there is only an implicit dependence on the interaction potential through the static structure factor. We now confine our discussion to charged systems. By introducing the normalized frequency $\Omega=\omega/\omega_{\mathrm{p}}$ with $\omega_{\mathrm{p}}$ the particle plasma frequency, employing the normalized wavenumber $q=kd$ and using $v_{\mathrm{T}}=\omega_{\mathrm{p}}\lambda_{\mathrm{D}}$ with $\lambda_{\mathrm{D}}^2=T/(4\pi{n}Q^2)$ the particle Debye length (not to be confused with the screening length $\lambda$), we obtain the \textbf{normalized SLFC dispersion relation}
\begin{align}
\Omega^2=\frac{q^2}{\Gamma}+\left[\frac{1}{S(q)}-1\right]\frac{q^2}{3\Gamma}\,.\label{dispersionrelationnormal}
\end{align}
This dispersion was derived in early works of Murillo\,\cite{theosl1,theosl2}, with \& without the Bohm-Gross term $q^2/\Gamma$ or $3k^2v_{\mathrm{T}}^2$. Note that this dispersion remains valid in the OCP limit ($\kappa\to0$) with the continuous transition from the acoustic-roton YOCP mode to the Langmuir-optical OCP mode facilitated by the removable singularity at $q=0$, as seen by the exact long wavelength limit of the OCP structure factor $S(q)=q^2/[(\widetilde{B}^{\mathrm{ex}}_{\mathrm{T}}+1){q}^2+3\Gamma]$ with $\widetilde{B}^{\mathrm{ex}}_{\mathrm{T}}$ the reduced excess isothermal bulk modulus. As expected, use of the RPA structure factor $S(q)=(q^2+\kappa^2)/(q^2+\kappa^2+3\Gamma)$, obtained for $G(\boldsymbol{k})\equiv0$ or $c(r)\equiv{c}(r\to\infty)=-\beta{U}(r)$ where $c(r)$ is the direct correlation function, leads to the standard dispersion relation in absence of correlations\,\cite{dusrev1}
\begin{align}
\Omega^2=q^2/(q^2+\kappa^2)+q^2/\Gamma\,.\label{dispersionrelationideal}
\end{align}

A similar expression has been obtained by Khrapak \& Cou{\"e}del based on physical intuition\,\cite{modulig}. These investigators constructed the dispersion relation by; inspecting the second frequency moment of the longitudinal current fluctuations (which corresponds to the fourth frequency moment rule for the dynamic structure factor\,\cite{slfccl1}) as well as the second frequency moment of the transverse current fluctuations, rewriting the potential contribution to these moments in terms of the generalized instantaneous elastic moduli\,\cite{moduli1}, conjecturing that the transverse mode cutoff at moderate coupling\,\cite{moduli2,moduli3,moduli4} allows the omission of the contribution of the shear modulus to the longitudinal modulus (which is then equal to the bulk modulus\,\cite{moduli5}) and adding the Bohm-Gross term artificially to ensure a proper transition to the weak coupling regime. Overall, in normalized units, their \textbf{semi-phenomenological QLCA-inspired dispersion relation} reads as\,\cite{modulig}
\begin{align}
\Omega^2=\frac{q^2}{\Gamma}+\widetilde{B}^{\mathrm{ex}}_{\infty}(q)\frac{q^2}{3\Gamma}\,,\label{dispersionrelationkhrapak}
\end{align}
where $\widetilde{B}^{\mathrm{ex}}_{\infty}(q)$ is the reduced excess part of the generalized instantaneous bulk modulus ${B}_{\infty}(q)$ that is given by\,\cite{moduli1}
\begin{align}
\widetilde{B}^{\mathrm{ex}}_{\infty}(q)&=\frac{\beta}{q^2}\int_0^{\infty}\left[J(qx)\frac{d^2{U}}{d^2x}+\frac{I(qx)}{x}\frac{d{U}}{dx}\right]x^2g(x)dx\,,\nonumber\\
J(y)&=-\frac{1}{3}-3\frac{\sin{y}}{y}+10\frac{\sin{y}}{y^3}-10\frac{\cos{y}}{y^2}\,,\nonumber\\
I(y)&=-\frac{2}{3}+4\frac{\sin{y}}{y}-10\frac{\sin{y}}{y^3}+10\frac{\cos{y}}{y^2}\,.
\end{align}
It suffices to show that $\widetilde{B}^{\mathrm{ex}}_{\infty}(q)\simeq{S}^{-1}(q)-1$ to prove that the dispersion relations Eqs.(\ref{dispersionrelationnormal},\ref{dispersionrelationkhrapak}) are almost identical. This can be anticipated from the long-wavelength limit $q\to0$, where $\widetilde{B}^{\mathrm{ex}}_{\infty}(q)$ collapses to the reduced excess part of the instantaneous bulk modulus $\widetilde{B}^{\mathrm{ex}}_{\infty}$ given by\,\cite{moduli8,moduli9}
\begin{align}
\widetilde{B}^{\mathrm{ex}}_{\infty}&=\frac{\beta}{6}\int_0^{\infty}\left[x\frac{d^2{U}}{d^2x}-2\frac{d{U}}{dx}\right]x^3g(x)dx\,,
\end{align}
and where the static structure factor $S(q)$ gets connected to $\widetilde{B}^{\mathrm{ex}}_{\mathrm{T}}$ via the thermodynamic fluctuation relation\,\cite{slfccl1}
\begin{align}
{S^{-1}(0)}-1&=\widetilde{B}^{\mathrm{ex}}_{\mathrm{T}}\,.
\end{align}
These two reduced excess bulk moduli are not identical quantities, but are expected to acquire similar values\,\cite{moduli0}.

\begin{figure}
	\centering
	\includegraphics[width=3.00in]{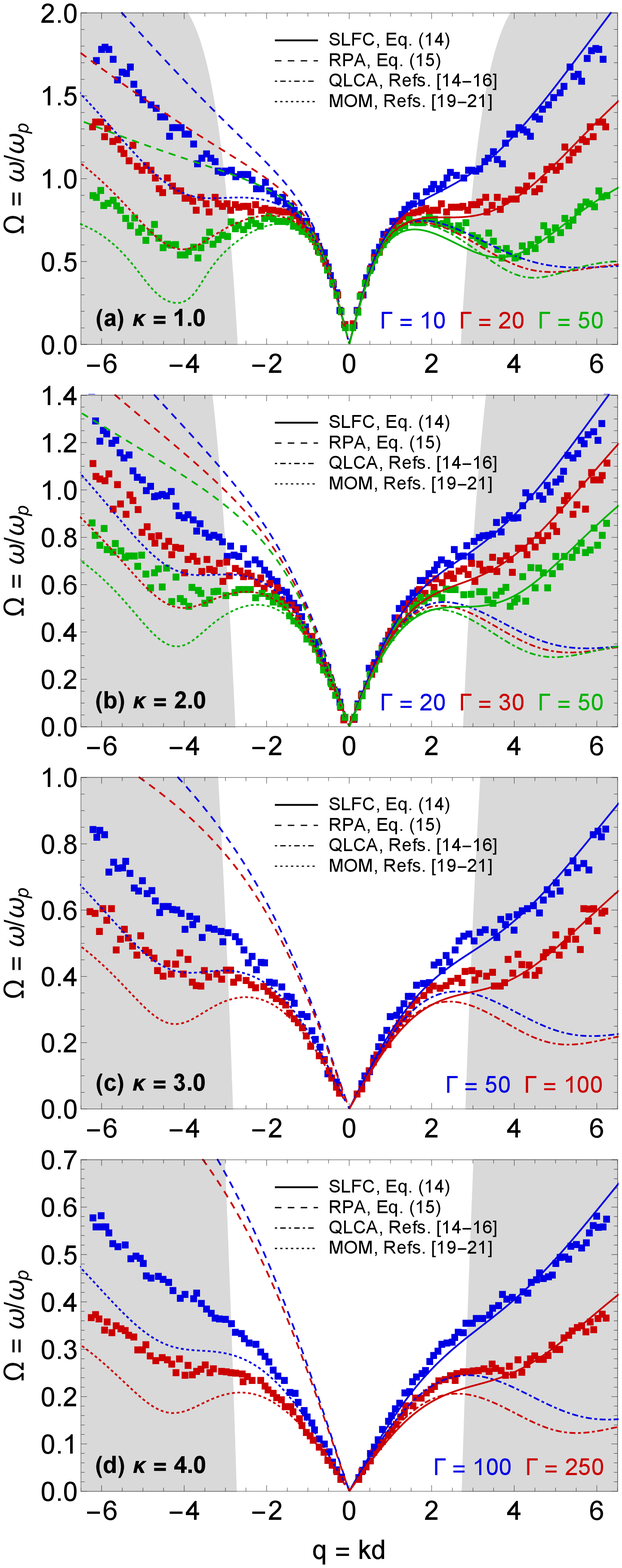}
	\caption{Dispersion relations for longitudinal waves in moderately coupled Yukawa liquids from the SLFC [Eq.(\ref{dispersionrelationnormal}), solid lines at $q\geq0$], the RPA [Eq.(\ref{dispersionrelationideal}), dashed lines at $q\leq0$], the QLCA (Refs.\cite{theoQL1,theoQL2,theoQL3}, dot-dashed lines at $q\geq0$), the MOM (Refs.\cite{theosr1,theosr2,theosr3}, dotted lines at $q\leq0$) and the MD simulations of Khrapak \& Cou{\"e}del\,\cite{modulig} (symbols at $\forall{q}$). Results for varying coupling parameters and (a) $\kappa=1$, (b) $\kappa=2$, (c) $\kappa=3$, (d) $\kappa=4$. In the shaded region, the analytic SLFC dispersion relation, based upon the asymptotic expansion of the plasma dispersion function, is not strictly valid. The structural input [$g(x),\,S(q)$] was computed with the VMHNC approach\,\cite{IETref5}.}\label{fig:dispersion_relation_v1}
\end{figure}

The approximate $\widetilde{B}^{\mathrm{ex}}_{\infty}(q)\simeq{S}^{-1}(q)-1$ and $\widetilde{B}^{\mathrm{ex}}_{\infty}\simeq\widetilde{B}^{\mathrm{ex}}_{\mathrm{T}}$ expressions have been validated by using accurate structural input for $g(r)$ and $S(q)$ from integral equation theory. In particular, the variational modified hypernetted-chain approach (VMHNC) based on the ansatz of bridge function quasi-universality\,\cite{IETref1,IETref2} and the empirically modified hypernetted-chain approach (IEMHNC) based on the near-isomorph invariance of Yukawa bridge functions\,\cite{IETref3,IETref4} have been employed. In a recent comparative study on Yukawa liquids, the IEMHNC and VMHNC approaches emerged as the most accurate integral equation theory methods with predictions of structural properties within $2\%$ of computer simulations inside the first coordination cell\,\cite{IETref5}. The VMHNC \& IEMHNC numerical results were nearly indistinguishable and verified that $\widetilde{B}^{\mathrm{ex}}_{\infty}\simeq\widetilde{B}^{\mathrm{ex}}_{\mathrm{T}}$ (see figure \ref{fig:long_wavelength}), $\widetilde{B}^{\mathrm{ex}}_{\infty}(q)\simeq{S}^{-1}(q)-1\,,\forall{q}\in[0,3]$ (see figure \ref{fig:any_wavelength}) at least for any $\kappa\in[0,4]$ and any $\Gamma/\Gamma_{\mathrm{m}}(\kappa)\in[0.05,1]$, where $\Gamma_{\mathrm{m}}(\kappa)$ is the YOCP melting line\,\cite{IETref6,IETref7}. It should be noted that the approximate equivalence degrades as the screening parameter increases and ceases to be valid beyond $\kappa=4$. However, such high $\kappa$ values are of less practical interest. It should also be noted that the approximate equivalence breaks down for roughly $q\gtrsim{3d}$. However, at higher wavenumbers, the longitudinal waves become anyway strongly suppressed by Landau damping.

Given the numerical verification of the approximate relation $\widetilde{B}^{\mathrm{ex}}_{\infty}(q)\simeq{S}^{-1}(q)-1$ in the regimes of interest, the dispersion relations described by Eqs.(\ref{dispersionrelationnormal},\ref{dispersionrelationkhrapak}) are nearly identical, which suggests that the analytic SLFC dispersion relation provides an accurate description of longitudinal modes in moderately coupled Yukawa liquids. This is clearly demonstrated in figure \ref{fig:dispersion_relation_v1}, where the predictions of the analytic SLFC dispersion relation are compared with MD-extracted dispersion relations for a number of YOCP state points\,\cite{modulig}. It should be noted that the analytic SLFC dispersion relation remains accurate even beyond its applicability range that originates from the use of the large argument expansion of the plasma dispersion function. The shaded region of the $(\Omega-q)$ plane roughly demarcates the frequency-wavenumber combinations for which the asymptotic result $[1+\zeta\mathcal{Z}(\zeta)]^{-1}=-2\zeta^2+3$ becomes inaccurate. The rather satisfactory accuracy of the analytic SLFC dispersion relation within the shaded region cannot be justified theoretically and might be illusory given the large uncertainties in MD extraction of real eigenfrequencies at short wavelengths\,\cite{MDwave1,modulig} due to the damping-induced broadening of the dynamic structure factor peak as $q$ increases\,\cite{theoQL3,MDwave5}. From figure \ref{fig:dispersion_relation_v1}, it is evident that the RPA dispersion relation, see Eq.(\ref{dispersionrelationideal}), exhibits pronounced deviations from MD results, confirming the important role of the static local field correction.

Figure \ref{fig:dispersion_relation_v1} also reveals that the SLFC dielectric formalism is more accurate than the prominent QLCA \& MOM approaches at moderate coupling. Loss of accuracy away from the strong coupling regime is a known QLCA shortcoming, since the approach explicitly neglects the direct thermal effects responsible for the Bohm-Gross term\,\cite{theoQL1,theoQL2,theoQL3}. The reduced accuracy of MOM in the moderate coupling regime is rather surprising and has to be attributed either to the approximation of the frequency-dependent Nevanlinna parameter function with its static value or to the ansatz that the dynamic structure factor exhibits an extremum at its static value\,\cite{theosr1,theosr2,theosr3}. Overall, the SLFC success for the moderately coupled YOCP most probably stems from the fact that the $\widetilde{B}^{\mathrm{ex}}_{\infty}(q)\simeq{S}^{-1}(q)-1$ relation implies that the fourth frequency moment rule is roughly satisfied (see the Khrapak \& Cou{\"e}del construction) without being imposed like the zero frequency moment rule.

To sum up, the SLFC dielectric formalism, earlier discussed by Ichimaru\,\cite{slfcmo2,slfcmo3} and Murillo\,\cite{theosl1,theosl2}, has been revisited in the context of moderately coupled YOCP liquids. A longitudinal dispersion relation has been derived that is nearly identical to a semi-phenomenological expression constructed by Khrapak \& Cou{\"e}del\,\cite{modulig} and that exhibits very good agreement with MD simulations. In the moderate coupling regime, the SLFC formalism has been demonstrated to be more accurate than established approaches that perform excellently at strong coupling.

The authors acknowledge the financial support of the Swedish National Space Agency under grant no.\,143/16.

The data that supports the findings of this study are available within the article.


\begin{thebibliography}{200}
\bibitem{dusrev1} V. E. Fortov, A. V. Ivlev, S. A. Khrapak, A. G. Khrapak and G. E. Morfill, \emph{Phys. Rep.} {\bf 421}, 1 (2005).
\bibitem{dusrev2} G. E. Morfill and A. V. Ivlev, \emph{Rev. Mod. Phys.} {\bf 81}, 1353 (2009).
\bibitem{dusrev3} M. Bonitz, C. Henning and D. Block, \emph{Rep. Prog. Phys.} {\bf 73}, 066501 (2010).
\bibitem{dusexp1} H. M. Thomas, M. Schwabe, M. Y. Pustylnik, C. A. Knapek, V. I. Molotkov, A. M. Lipaev, O. F. Petrov, V. E. Fortov and S. A. Khrapak \emph{Plasma Phys. Control. Fusion} {\bf 61} 014004 (2019).
\bibitem{dusexp2} H. Boroudjerdi, Y.-W. Kim, A. Naji, R. R. Netz, X. Schlagberger and A. Serr, \emph{Phys. Rep.} {\bf 416}, 129 (2005).
\bibitem{dusexp3} T. C. Killian, T. Pattard, T. Pohl and J. M. Rost, \emph{Phys. Rep.} {\bf 449}, 77 (2007).
\bibitem{theoGH1} M. A. Berkovsky, \emph{Phys. Lett. A} {\bf 166}, 365 (1992).
\bibitem{theoGH2} P. K. Kaw and A. Sen, \emph{Phys. Plasmas} {\bf 5}, 3552 (1998).
\bibitem{theoGH3} P. K. Kaw, \emph{Phys. Plasmas} {\bf 8}, 1870 (2001).
\bibitem{theoVD1} A. Diaw and M. S. Murillo, \emph{Phys. Rev. E} {\bf 92}, 013107 (2015).
\bibitem{theoVD2} G. Dharuman, L. G. Stanton, J. N. Glosli and M. S. Murillo, \emph{J. Chem. Phys.} {\bf 146}, 024112 (2017).
\bibitem{theome1} M. S. Murillo, \emph{Phys. Rev. Lett.} {\bf 85}, 2514 (2000).
\bibitem{theome2} J. P. Mithen, J. Daligault, B. J. B. Crowley and G. Gregori, \emph{Phys. Rev. E} {\bf 84}, 046401 (2011).
\bibitem{theoQL1} G. Kalman and K. I. Golden, \emph{Phys. Rev. A} {\bf 41}, 5516 (1990).
\bibitem{theoQL2} K. I. Golden and G. J. Kalman, \emph{Phys. Plasmas} {\bf 7}, 14 (2000).
\bibitem{theoQL3} Z. Donko, G. J. Kalman and P. Hartmann, \emph{J. Phys.: Condens. Matter} {\bf 20}, 413101 (2008).
\bibitem{theosl1} M. S. Murillo, \emph{Phys. Plasmas} {\bf 5}, 3116 (1998).
\bibitem{theosl2} M. S. Murillo, \emph{Phys. Plasmas} {\bf 7}, 33 (2000).
\bibitem{theosr1} Yu. V. Arkhipov, A. Askaruly, A. E. Davletov, D. Yu. Dubovtsev, Z. Donko, P. Hartmann, I. Korolov, L. Conde and I. M. Tkachenko, \emph{Phys. Rev. Lett.} {\bf 119}, 045001 (2017).
\bibitem{theosr2} Yu. V. Arkhipov, A. B. Ashikbayeva, A. Askaruly, M. Bonitz, L. Conde, A.E. Davletov, T. Dornheim, D. Yu. Dubovtsev, S. Groth, Kh. Santybayev, S.A. Syzganbayeva and I.M Tkachenko, \emph{Contrib. Plasma Phys.} {\bf 58}, 967 (2018).
\bibitem{theosr3} Yu. V. Arkhipov, A. Ashikbayeva, A. Askaruly, A. E. Davletov, D. Yu. Dubovtsev, Kh. S. Santybayev, S. A. Syzganbayeva, L. Conde and I. M. Tkachenko, \emph{Phys. Rev. E} {\bf 102}, 053215 (2020)
\bibitem{theocs1} S. A. Khrapak and H. M. Thomas, \emph{Phys. Rev. E} {\bf 91}, 033110 (2015).
\bibitem{theocs2} S. A. Khrapak, B. Klumov, L. Cou{\"e}del and H. M. Thomas, \emph{Phys. Plasmas} {\bf 23}, 023702 (2016).
\bibitem{theocs3} L. G. Silvestri, G. J. Kalman, Z. Donko, P. Hartmann, M. Rosenberg, K. I. Golden and S. Kyrkos, \emph{Phys. Rev. E} {\bf 100}, 063206 (2019).
\bibitem{theocs4} S. A. Khrapak, \emph{Phys. Plasmas} {\bf 26}, 103703 (2019).
\bibitem{MDwave1} H. Ohta and S. Hamaguchi, \emph{Phys. Rev. Lett.} {\bf 84}, 6026 (2000).
\bibitem{MDwave2} S. Hamaguchi and H. Ohta, \emph{Phys. Scr.} {\bf T89}, 127 (2001).
\bibitem{MDwave3} J. Goree, Z. Donko and P. Hartmann, \emph{Phys. Rev. E} {\bf 85}, 066401 (2012).
\bibitem{MDwave4} J. P. Mithen, \emph{Phys. Rev. E} {\bf 89}, 013101 (2014).
\bibitem{MDwave5} Y. Choi, G. Dharuman and M. S. Murillo, \emph{Phys. Rev. E} {\bf 100}, 013206 (2019).
\bibitem{modulig} S. Khrapak and L. Cou{\"e}del, \emph{Phys. Rev. E} {\bf 102}, 033207 (2020).
\bibitem{frenkel} V. V. Brazhkin, A. G. Lyapin, V. N. Ryzhov, K. Trachenko, Yu. D. Fomin and E. N. Tsiok, \emph{Phys. Usp.} {\bf 55}, 1061 (2012).
\bibitem{kirkwoo} R. J. F. Leote de Carvalho, R. Evans and Y. Rosenfeld, \emph{Phys. Rev. E} {\bf 59}, 1435 (1999).
\bibitem{slfcge1} S. Ichimaru, \emph{Rev. Mod. Phys.} {\bf 54}, 1017 (1982).
\bibitem{slfcge2} S. Ichimaru, H. Iyetomi and S. Tanaka, \emph{Phys. Rep.} {\bf 149}, 91 (1987).
\bibitem{slfcge3} G. Giuliani and G. Vignale, \emph{Quantum theory of the electron liquid}, (Cambridge University Press, New York, 2005).
\bibitem{slfcge4} K. S. Singwi and M. P. Tosi, \emph{Solid State Phys.} {\bf 36}, 177 (1981).
\bibitem{slfcge5} T. Dornheim, S. Groth and M. Bonitz, \emph{Phys. Rep.} {\bf 744}, 1 (2018).
\bibitem{slfcqu1} K. S. Singwi, M. P. Tosi, R. H. Land and A. Sj{\"o}lander, \emph{Phys. Rev.} {\bf 176}, 589 (1968).
\bibitem{slfcqu2} S. Tanaka and S. Ichimaru, \emph{Phys. Rev. B} {\bf 39}, 1036 (1989).
\bibitem{slfcqu3} S. Tanaka, \emph{J. Chem. Phys.} {\bf 145}, 214104 (2016).
\bibitem{slfccl1} J.-P. Hansen and I. R. McDonald, \emph{Theory of simple liquids}, (Academic Press, New York, 2005).
\bibitem{slfccl2} J. P. Boon and S. Yip, \emph{Molecular hydrodynamics}, (Dover Publications, New York, 1991).
\bibitem{slfccl3} A. A. Kugler, \emph{J. Stat. Phys.} {\bf 8}, 107 (1973).
\bibitem{slfcmo1} B. D. Fried and S. D. Conte, \emph{The plasma dispersion function}, (Academic Press, New York, 1961).
\bibitem{slfcmo2} K. Tago, K. Utsumi and S. Ichimaru, \emph{Prog. Theor. Phys.} {\bf 65}, 54 (1981).
\bibitem{slfcmo3} S. Ichimaru, \emph{Statistical plasma physics volume I: basic principles}, (CRC Press, New York, 1991).
\bibitem{moduli1} R. Nossal, \emph{Phys. Rev.} {\bf 166}, 81 (1968).
\bibitem{moduli2} M. S. Murillo, \emph{Phys. Rev. Lett.} {\bf 85}, 2514 (2000).
\bibitem{moduli3} S. A. Khrapak, A. G. Khrapak, N. P. Kryuchkov and S. O. Yurchenko, \emph{J. Chem. Phys.} {\bf 150}, 104503 (2019).
\bibitem{moduli4} V. V. Brazhkin, Y. D. Fomin, A. G. Lyapin, V. N. Ryzhov and K. Trachenko, \emph{Phys. Rev. E} {\bf 85}, 031203 (2012).
\bibitem{moduli5} R. Zwanzig, \emph{Phys. Rev.} {\bf 156}, 190 (1967).
\bibitem{moduli8} N. K. Ailawadi, \emph{Phys. Rep.} {\bf 57}, 241 (1980).
\bibitem{moduli9} S. Khrapak, \emph{Phys. Rev. E} {\bf 100}, 032138 (2019).
\bibitem{moduli0} P. Schofield, \emph{Proc. Phys. Soc.} {\bf 88}, 149 (1966).
\bibitem{IETref1} Y. Rosenfeld, \emph{J. Stat. Phys.} {\bf 42}, 437 (1986).
\bibitem{IETref2} G. Faussurier, \emph{Phys. Rev. E} {\bf 69}, 066402 (2004).
\bibitem{IETref3} P. Tolias and F. Lucco Castello, \emph{Phys. Plasmas} {\bf 26}, 043703 (2019).
\bibitem{IETref4} F. Lucco Castello, P. Tolias and J. C. Dyre, \emph{J. Chem. Phys.} {\bf 154}, 034501 (2021).
\bibitem{IETref5} F. Lucco Castello and P. Tolias, \emph{Contrib. Plasma Phys.} {\bf 61}, e202000105 (2021).
\bibitem{IETref6} S. Hamaguchi, R. T. Farouki and D. H. E. Dubin, \emph{Phys. Rev. E} {\bf 56}, 4671 (1997).
\bibitem{IETref7} O. Vaulina, S. Khrapak and G. Morfill, \emph{Phys. Rev. E} {\bf 66}, 016404 (2002).
\end{thebibliography}
\end{document}